\documentclass[
amsmath, %
floatfix, %
twocolumn, %
reprint, %
prl, %
aps, %
]{revtex4-2}

\renewcommand{\thispagestyle}[1]{}
\usepackage[T1]{fontenc}
\usepackage[utf8]{inputenc}
\usepackage{graphicx}
\usepackage{latexsym}
\usepackage{amsthm}
\usepackage{color}
\usepackage{mathtools}
\usepackage{xcolor}
\usepackage[section]{placeins}

\usepackage[]{newtxtext}
\usepackage[nosymbolsc,smallerops,bigdelims]{newtxmath}
\DeclareMathAlphabet{\mathcal}{OMS}{cmsy}{m}{n}

\DeclareMathAlphabet{\mathcalb}{OMS}{cmsy}{b}{n}
\usepackage{bm}
\usepackage{hyperref}
\hypersetup{
	colorlinks,
	linkcolor={blue!90!black},
	citecolor={blue!90!black},
	urlcolor={blue!90!black}
}
\makeatletter
\renewcommand*{\eqref}[1]{%
	\hyperref[{#1}]{\textup{\tagform@{\ref*{#1}}}}%
}
\makeatother

\DeclarePairedDelimiter\lr{\lparen}{\rparen}
\DeclarePairedDelimiter\Lr{\lbrack}{\rbrack}

\DeclarePairedDelimiter\abs{\lvert}{\rvert}
\DeclarePairedDelimiter\avg{\langle}{\rangle}

\DeclarePairedDelimiter\ceil{\lceil}{\rceil}
\DeclarePairedDelimiter\avgy{\lbrack}{\rbrack_{y}}

\DeclarePairedDelimiterX{\comm}[2]{\lbrack}{\rbrack}{#1, #2}
\DeclarePairedDelimiterX{\acomm}[2]{\lbrace}{\rbrace}{#1, #2}

\DeclarePairedDelimiterX{\braket}[2]{\langle}{\rangle}{#1\delimsize\vert #2}
\DeclarePairedDelimiterX{\ketbra}[2]{\rvert}{\lvert}{#1 \delimsize\rangle\!\delimsize\langle #2}
\DeclarePairedDelimiterX{\matrixel}[3]{\langle}{\rangle}{#1 \delimsize\rvert #2 \delimsize\lvert #3}

\newcommand{\Tr}{\mathrm{Tr}}
\newcommand{\lkaka}{\lambda_{k\alpha k'\alpha'}}
\newcommand{\Skaka}{\sum_{k\alpha k'\alpha'}}
\newcommand{\Ska}{\sum_{k\alpha}}
\newcommand{\pka}{p_{\alpha}^{(k)}}
\newcommand{\qka}{q_{\alpha}^{(k)}}
\newcommand{\pkkaa}{p_{\alpha'}^{(k')}}
\newcommand{\qkkaa}{q_{\alpha'}^{(k')}}
\newcommand{\Mpq}{\mathcal{M}_{pq}}
\newcommand{\JBM}{\frac{J\beta}{M}}
\newcommand{\pkia}{p_{i\alpha}^{(k)}}
\newcommand{\qkia}{q_{i\alpha}^{(k)}}
\newcommand{\Rkk}{\mathcal{R}_{kk'}}
\newcommand{\Ukk}{\mathcal{U}_{kk'}}

\newcommand{\Qaa}{\mathcal{Q}_{\alpha\alpha'}}
\newcommand{\QEA}{\mathcal{Q}_\mathrm{EA}}

\newcommand{\DD}{\int\!\!\mathrm{D}}
\newcommand{\qo}{q_{0}}
\newcommand{\qi}{q_{1}}
\newcommand{\uo}{u_{0}}
\newcommand{\ui}{u_{1}}
\newcommand{\Tc}{T_{\mathrm{c}}}
\newcommand{\kB}{k_{\mathrm{B}}}

\newcommand{\ii}{\mathrm{i}}
\newcommand{\e}{\mathrm{e}}

\newcommand{\intid}{\int\limits_{-\infty}^{\infty}\!\!\mathrm{d}}

\begin{document}
\title{Physical manifestation of replica symmetry breaking
in a quantum glass of bosons with off-diagonal disorder}
\author{Anna M. Piekarska}
\email{a.piekarska@intibs.pl}
\affiliation{Institute of Low Temperature and Structure Research, Polish Academy of Sciences, 
Ok\'{o}lna 2, 50-422 Wroc\l{}aw, Poland}
\author{Tadeusz K. Kope\'{c}}
\affiliation{Institute of Low Temperature and Structure Research, Polish Academy of Sciences, 
Ok\'{o}lna 2, 50-422 Wroc\l{}aw, Poland}
\begin{abstract}
Glassiness occurs when disorder and frustration cause local degrees of freedom to freeze despite the lack of long-range order.
In systems of interacting bosons, such glassiness may involve a purely quantum degree of freedom---local phases of particle wave functions---partly analogous to spins in spin glasses.
However, experimental identification of such phases is difficult because it requires prohibitively long measurement times
or recourse to the elusive Edwards-Anderson order parameter.
Moreover, the off-diagonal character of the phase makes it seemingly even harder to capture via typical observables.
To address this issue, we study a system of strongly interacting bosons with random hoppings
that features off-diagonal glassiness exhibiting replica symmetry breaking (RSB).
We find that the glass phase is compressible,
which distinguishes it from the Mott insulator.
Thus, we establish a direct correspondence between phase-based glassy order and a measurable density-based thermodynamic observable.
We use a framework adopted from spin glasses, including the replica trick within the one-step RSB scheme, to obtain meaningful results in the glass phase and to characterize the order parameters, RSB structure, slow relaxation, and compressibility. 
Glassiness in particle systems could thus be experimentally identified via measurements of compressibility, such as probing density fluctuations or the particle-number response to a trapping potential.
\end{abstract}
\maketitle
%
{\it Introduction.}---
Glasses with quenched disorder, particularly quantum glasses, have recently seen a revival of interest across various directions.
The original classical~\cite{LacroixAChezToine2024_JSP191,Yeo2023_PRE108,Paga2021_JSM2021,Moore2018_PRL120}
and quantum spin-glass~\cite{Tikhanovskaya2024_PQ5,Kavokine2024_PRL133,Baldwin2020_PRX10} problems
are receiving new solutions~\cite{book_Charbonneau2023} and recognition~\cite{Parisi2023_RMP95}.
New research branches are also emerging, uncovering glassy physics in fields such as
random lasers~\cite{Gradenigo2020_PRR2,Ghofraniha2015_NC6},
boson glasses~\cite{Piekarska2022_PRB105,Piekarska2018_PRL120},
or superglasses~\cite{Biroli2008_PRB78,Tam2010_PRL104}.
Finally, the rise of quantum simulation is opening new avenues
for revisiting glasses experimentally~\cite{Kroeze2025_S389,Marsh2024_PRX14,Harris2018_S361,Rotondo2015_PRB91}.
Quenched off-diagonal disorder in such systems, i.e., static randomness in couplings that frustrates the studied degrees of freedom,
leads to a rugged free energy landscape with multiple metastable quasi-equilibria.
There, relaxation requires thermal fluctuations or tunneling through energy barriers, making it extremely slow and dependent on the initial state.
Due to this, both theoretical descriptions and experimental probing of such systems are notoriously difficult.

The progress in understanding spin-glass physics has been marked by three milestones:
introduction of the Edwards-Anderson (EA) order parameter~\cite{Edwards1975_JPFMP5},
the Sherrington-Kirkpatrick (SK) model~\cite{Sherrington1975_PRL35} that provided a simple description of the phenomenon,
and the replica symmetry breaking (RSB) scheme~\cite{Parisi1979_PLA73,Parisi1979_PRL43} that fully solved the SK model.
The earlier replica-symmetric (RS) approach provided a correct description of the disordered phase but broke down in describing the glass.
The RSB construction, apart from fixing this issue, has also broadened the understanding of glassiness,
providing a physical interpretation of the replicas~\cite{Mezard1984_PRL52},
which can be thought of as hierarchically organized metastable states.
Despite the SK model being solved by RSB,
the physical implications remain an area of active research~\cite{book_Charbonneau2023}.
Moreover, due to the complexity of RSB, its application to other models is not straightforward 
and requires simplifications or substantial computational effort~\cite{book_Charbonneau2023}.

Glasses are an example of a complex system,
i.e., a system characterized by nonlinearity, emergence, spontaneous order, or feedback loops~\cite{book_Stein2013}.
Similarities between various types of complex systems give hope that a breakthrough in one could advance others, 
even in distant fields.
For example, researching glasses might help in understanding protein folding,
which, like glasses, is governed by rugged free-energy landscapes and exhibits the slow dynamics resulting from them~\cite{Galpern2024_PNAS121}.
Another example of such a system of current interest is neural networks,
with their underlying minimization problem directly mapping to the spin glass Hamiltonian~\cite{Choromanska2015}.
Interest in glassy physics has spiked again after Parisi received the 2021 Nobel prize~\cite{Parisi2023_RMP95}.
This recent surge can leverage the enormous progress in computational and experimental capabilities compared to the 1980s.
For example, the rise of quantum simulation~\cite{Georgescu2014_RMP86} has provided the means to
study models that were previously inaccessible to computational treatment.
Another direct and new realization of glassy phases can be found in random lasers,
where random scattering of light leads to a glassy state of photons,
facilitating the observation of RSB~\cite{Ghofraniha2015_NC6}.

Although the majority of the theoretical work on glasses has concerned the classical case,
the influence of quantum effects cannot be neglected,
as they lead to significant renormalization of critical temperatures and relaxation times by allowing tunneling,
and enable quantum phase transitions~\cite{Wu1991_PRL67,Bray1980_JPC13,Usadel1988_NPB5}.
However, the quantum case has proven to be much more challenging, often requiring heavy numerical support
due to non-commuting Hamiltonian terms.
With current computational resources,
they regain attention as previously intractable problems become accessible.
Apart from studying quantum counterparts of classical models,
gaining insights into purely quantum systems without classical counterparts is no less important.
In this context, the search for novel glassy phases involving freezing in previously unexplored degrees of freedom
aligns with the broader research frontier on nontrivial correlated phases of matter.

Among the testbeds for discovering novel phases, quantum simulators play a significant role as they
pave the way for shaping system potential energy or arranging sites in arbitrary configurations. This allows realizing various models,
from basic solid-state theory all the way to exotic artificial models or known models in otherwise unphysical parameter regimes~\cite{Nataf2010_NC1}.
For example, in optical lattices,
precisely arranged laser beams create lattices of chosen dimensionality~\cite{Jaksch1998_PRL81} resembling solids. They can be populated with both fermionic and bosonic atomic isotopes, allowing the study of a wide range of models. 
Such systems feature tunable interactions that extend to the regime of strong correlations, offering insights into the superfluid-Mott insulator phase transition~\cite{Greiner2002_N415}.
Quantum simulators not only control interactions, energy scales, and system geometry,
but also facilitate the introduction of disorder into the system~\cite{Fallani2008}.
In this way, one can study a system of (bosonic) particles subjected to disorder in a controlled manner.

This setting has produced the Bose glass,
an amorphous state of bosons in which diagonal site disorder leads to an insulating state, with localization preventing superfluidity.
It has been widely studied in the context of dirty bosons~\cite{Fisher1989_PRB40},
and later revisited thanks to quantum simulator platforms~\cite{Gimperlein2005_PRL95}.
While it exhibits glassy features such as slow dynamics, it is not generally considered a glass in the strict Parisi-type sense used here.
Although a formulation introducing an EA-like order parameter defined in terms of density fluctuations exists~\cite{Thomson2014_EL108}, it captures persistent correlations rather than frozen local degrees of freedom,
in strong contrast to the glassiness considered here. The fundamental difference stems from the diagonal disorder in the Bose glass, as opposed to the off-diagonal disorder considered here.
This distinction is similar to that between random-field magnetic systems~\cite{Imry1975_PRL35} and random-bond systems---spin glasses.

Glassiness in bosonic systems is appealing, 
as it offers rich physical intuitions and a well-established experimental setting.
As expected for systems analogous to spin glasses,
their treatment is challenging, leaving them less explored.
This connection, however, promises the opportunity to integrate the rich physics of quenched disorder, frustration, and RSB
into the realm of bosons, where these can be confronted with interactions and superfluidity~\cite{Piekarska2022_PRB105},
and where the complex RSB structure can be explored not merely as a theoretical concept, but as a physical reality within the reach of current experiments.
The resulting combination could be simulated in an optical lattice.
A setup for measuring overlap functions, a building block of the EA order parameter,
has been proposed~\cite{Morrison2008_NJP10} even before such a bosonic glass with off-diagonal disorder was considered~\cite{Piekarska2018_PRL120}.
The system has thus far been studied in the RS approximation only, insufficient to properly describe the glass phase,
and therefore requires further examination via RSB.

In this work, we consider a system of strongly interacting bosons with random tunneling amplitudes that can be
realized in an optical lattice and features a glass phase with RSB in which bosonic phases are frozen.
We describe the system using the Bose-Hubbard model
with the hopping term being a random variable with zero mean.
This choice allows us to focus solely on the glassy physics without accounting for the competition from superfluidity \cite{Piekarska2022_PRB105}.
The latter cannot emerge in this setting, as it 
requires breaking the $U(1)$ symmetry through the mean hopping amplitude.

The glass we consider can be characterized using an EA order parameter $\QEA=[|\avg{a_i}|^2]$, with $a_i$ representing the annihilation of a boson in site $i$ in the systems, while $\avg{\cdot}$ and $[\cdot]$ denote thermodynamic and disorder averages, respectively. It captures the freezing of local boson phases, a natural analog of the spin-glass order parameter with connections to the concept of phase glasses~\cite{Kopec1995_PRB52a,Dalidovich2002_PRL89,Wu2006_PRB73}.

We solve the model using a scheme based on the SK approach to spin glasses~\cite{Sherrington1975_PRL35},
which can be extended to off-diagonal quantum glasses of bosons~\cite{Piekarska2018_PRL120}.
Starting from the disorder-averaged (quenched) free energy of the system,
we employ the replica trick to make the averaging feasible and obtain a replicated system allowing for RSB emergence. From the saddle-point solution in the thermodynamic limit,
we recover a set of self-consistent equations describing an effective classical problem.
We solve these equations numerically within the one-step replica symmetry breaking (1-RSB) approximation.

We observe the phase transition between the glass and disordered phases.
In characterizing the phases,
we focus particularly on the physics of the unstable glass phase.
We analyze the order parameters and RSB structure, and find long relaxation times.
Our central result is that the glass phase is compressible.
This leads to a striking realization that a zero-mean random tunneling, primarily affecting the phases of wave functions,
results in a physically distinct response in particle density
compared to that of the Mott insulator, the state of the system under the same parameters in the absence of randomness.

{\it Model.}---
We describe the system using a variant of the Bose-Hubbard Hamiltonian~\cite{Fisher1989_PRB40} adapted to our problem,
\begin{equation}
	H = \sum_{i<j}J_{ij}\lr*{a_i^\dagger a_j+a_j^\dagger a_i}
	    + \frac{U}{2}\sum_{i} \hat{n}_i\lr*{\hat{n}_i-1} - \mu \sum_{i} \hat{n}_i,
\end{equation}
where $i = 1\ldots N$ numbers sites,
and $\hat{n}_i=a_i^\dagger a_i$.
The first term describes hopping between sites $i$ and $j$,
with a Gaussian-distributed zero-mean hopping integral $J_{ij}$ with variance $J/N$.
The second term represents the on-site interaction,
and the third accounts for the chemical potential.
In derivations, we switch to the quasi-momentum and quasi-position operators
$\hat{P} = \ii(a^{\dagger}-a)/\sqrt{2}$, $\hat{Q} = (a^{\dagger}+a)/\sqrt{2}$
with eigenvalues $p$, $q$.

To solve this model,
we follow the SK scheme~\cite{Sherrington1975_PRL35}, known from spin-glass systems,
adapted to quantum systems~\cite{Bray1980_JPC13,Usadel1988_NPB5} and subsequently to bosonic quantum systems~\cite{Piekarska2018_PRL120}.
We start with the free energy averaged over quenched disorder $[F] = -(1/\beta) [\ln \Tr \exp(-\beta H)]$.
The replica trick, $[F]=\lim_{n\to0}([Z^n]-1)/n$~\cite{Edwards1975_JPFMP5}, facilitates the averaging and introduces replicas labeled by an index $\alpha=1\ldots n$,
with the replicated Hamiltonian $H_{\mathrm{repl}}=H_P+H_Q+H_n,$, where $H_P = \sum_\alpha\sum_{i < j} J_{ij}\hat{P}_{i\alpha}\hat{P}_{j\alpha}$, $H_Q = \sum_\alpha\sum_{i < j} J_{ij}\hat{Q}_{i\alpha}\hat{Q}_{j\alpha}$, $H_n = (U/2)\sum_{i\alpha}\hat{n}_{i\alpha}^2-(\mu+U/2)\sum_{i\alpha}\hat{n}_{i\alpha}$.
To handle the quantum nature of the Hamiltonian, we apply the Trotter-Suzuki formula~\cite{Suzuki1976_CMP51}, which introduces an index $k=1\ldots M$.
Finally, we perform Gaussian averaging, take the thermodynamic limit, and apply the saddle point method to find the effective free energy.
A detailed derivation of the above steps can be found in Ref.~\onlinecite{Piekarska2022_PRB105}.
At this point, the effective free energy reads
\begin{equation}\label{eq:frene}
	\mathcal{F} = \Skaka\Lr*{2\lkaka^{2} + \frac{1}{2}\lr*{\lkaka^{PQ}}^{2}} - \ln \Tr \exp(-\beta \mathcal{H}),
\end{equation}
with an effective Hamiltonian
\begin{multline}\label{eq:effham}
	-\beta \mathcal{H} = \JBM\Skaka\Lr[\Bigg]{
		\lkaka\lr*{\pka \pkkaa + \qka \qkkaa}\\
		+ \lkaka^{PQ}\pka \qkkaa
	}
	+\Mpq,
\end{multline}
where $\mathcal{M}_{pq}=\prod_i \mathcal{M}_{pq}^{(i)}$ is a product of matrix elements,
\begin{equation}
	\mathcal{M}_{pq}^{(i)}
	=\prod_{k\alpha}
		\matrixel*{\pkia}{e^{-\frac{\beta H_n}{2M}}}{\qkia}
		\matrixel*{\qkia}{e^{-\frac{\beta H_n}{2M}}}{p_{i\alpha}^{(k+1)}},
\end{equation}
while $\lkaka$ and $\lkaka^{PQ}$ are variables introduced in the saddle point method,
defined self-consistently as
\begin{equation}
    \!\!\lkaka = \frac{J\beta}{2M}\avg*{\pka \pkkaa},\quad
	\lkaka^{PQ} = \JBM\avg*{\pka \qkkaa},
\end{equation}
with thermal averages taken with the effective Hamiltonian,
$\avg{\mathcal{A}} = (1/Z) \Tr \mathcal{A} \exp(-\beta \mathcal{H})$.

The variables $\lkaka$ and $\lkaka^{PQ}$
have a different physical meaning depending on whether $\alpha = \alpha'$ or not.
Thus, we explicitly split them, defining new variables
\begin{subequations}
\begin{eqnarray}\label{eq:decomp-a}
	\lkaka &=& \frac{J\beta}{2M}\Lr[\Big]{\Rkk\delta_{\alpha \alpha'}
		+ \lr{1-\delta_{\alpha \alpha'}}\Qaa},\\\label{eq:decomp-b}
	\lkaka^{PQ} &=& \JBM\Lr[\Big]{\Ukk\delta_{\alpha \alpha'}
		+ \lr{1-\delta_{\alpha \alpha'}}\mathcal{U}_{\alpha\alpha'}}.
\end{eqnarray}
\end{subequations}
Here, $\Rkk$ are the dynamic self-interactions, diagonal in $\alpha$ and dependent on $\abs{k-k'}$,
while $\Qaa$ are defined for $\alpha\neq\alpha'$ and independent of $k$ and $k'$.
$\Ukk$ and $\mathcal{U}_{\alpha\alpha'}$ are defined analogously.
To take the limit $n\to 0$ and recover the quenched free energy, an assumption about the $\Qaa$ matrix is necessary.
The simplest one is the replica-symmetric approximation, in which $\Qaa = q$,
previously considered in this system \cite{Piekarska2022_PRB105}.
Here, we go beyond this approximation and explore the 1-RSB solution,
which should yield more accurate results in the glassy phase
at the cost of increased complexity in the derivation and of close-to-prohibitively expensive numerical calculations.
In the 1-RSB approximation, we assume
\begin{equation}
	\Qaa = \begin{cases}
		0     & \mbox{for }   \alpha = \alpha',\\
		\qi   & \mbox{for }   \ceil{\frac{\alpha}{m}} = \ceil{\frac{\alpha'}{m}},\\
		\qo   & \mbox{otherwise},
	\end{cases}
\end{equation}
where $m$ is the parameter specifying the block size in replica space, indicating a cluster in the configuration space, while $\qi$ and $\qo$ are intra- and inter-cluster overlaps, respectively.
Treating $\mathcal{U}_{\alpha\alpha'}$ analogously, we introduce variables $\uo$, $\ui$.

We handle the free energy and self-consistent equations within a standard 1-RSB scheme
(see Appendix~A for details),
and arrive at
\begin{align}
	\mathcal{F} ={}& \frac{1}{2}\lr*{\JBM}^{2}\sum_{kk'}\lr*{\Rkk^{2}+\Ukk^{2}}- \frac{1}{m}\DD z \ln \DD y Z^{m} \nonumber\\
	&{}+ \frac{1}{2}\lr*{J\beta}^{2}\Lr*{(m-1)\qi^{2}-m\qo^{2}+(m-1)\ui^{2}-m\uo^{2}},
\end{align}
where we have defined
\begin{equation}
	\DD y \equiv \iiint\limits_{-\infty}^{\infty} \mathrm{d}y_{P}\mathrm{d}y_{Q}\mathrm{d}y_{B}
		\exp\lr*{-y_{P}^{2}-y_{Q}^{2}-y_{B}^{2}},
\end{equation}
and $y_P$ ($y_Q$; $y_B$) is a field coupled to terms involving $\sum_\alpha P_\alpha$ ($\sum_\alpha Q_\alpha$; $\sum_\alpha P_\alpha+Q_\alpha$) introduced in the Hubbard-Stratonovich transformation~\cite{book_Sachdev2023},
and analogously for $z$.

The final form of the effective Hamiltonian reads
\begin{multline}\label{eq:finalham}
	\!\!\!\!\!\!-\beta \mathcal{H} = \JBM \!\sum_{k} \Lr[\Bigg]{
		\sqrt{2(\qo{-}\uo)}\lr*{z_{P}p_{k}{+}z_{Q}q_{k}}{+}\sqrt{2\uo}z_{S}(p_{k}{+}q_{k})\\
		+\sqrt{2(\qo-\uo)}\lr*{z_{P}p_{k}+z_{Q}q_{k}}+\sqrt{2\uo}z_{S}(p_{k}+q_{k})}\\
	+\frac{1}{2}\lr*{\JBM}^{2}\sum_{kk'}\Lr[\Bigg]{\lr*{\Rkk-\qi}\lr*{p_{k}p_{k'}+q_{k}q_{k'}}\\
		+2\lr*{\Ukk-\ui}p_{k}q_{k'}},
\end{multline}
and the self-consistent equations are
\begin{subequations}
\begin{eqnarray}\label{eq:selfcon-a}
	\Rkk &=& \DD z \avgy*{\avg*{p_{k}p_{k'}}},\\
	\Ukk &=& \DD z \avgy*{\avg*{p_{k}q_{k'}}},\\
	\qi  &=& \frac{1}{M^{2}}\sum_{kk'}\DD z \avgy*{\avg{p_{k}}\avg{p_{k'}}},\\
	\qo  &=& \frac{1}{M^{2}}\sum_{kk'}\DD z \avgy*{\avg{p_{k}}}\avgy*{\avg{p_{k'}}},\\
	\ui  &=& \frac{1}{M^{2}}\sum_{kk'}\DD z \avgy*{\avg{p_{k}}\avg{q_{k'}}},\\\label{eq:selfcon-f}
	\uo  &=& \frac{1}{M^{2}}\sum_{kk'}\DD z \avgy*{\avg{p_{k}}}\avgy*{\avg{q_{k'}}},
\end{eqnarray}
\end{subequations}
where we have defined
\begin{equation}
	\avgy*{\mathcal{A}} \equiv \frac{\DD y \mathcal{A} Z^{m}}{\DD y Z^{m}}.
\end{equation}
Finally, we solve them numerically, as outlined in Appendix~B.

\begin{figure}[htb]
	\includegraphics[width=\columnwidth]{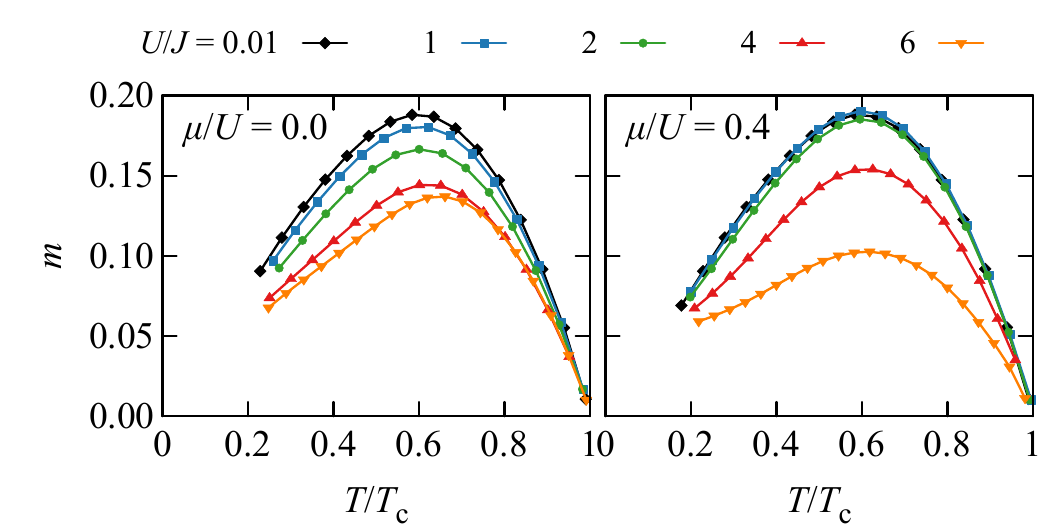}
	\caption{Temperature dependence of parameter $m$
		at various values of the chemical potential $\mu/U$ (panels) and inverse disorder $U/J$ (point shapes),
		as marked on the plots.
		Lines are to guide the eye only.}
	\label{fig:m}
\end{figure}

\begin{figure}[htb]
	\includegraphics[width=\columnwidth]{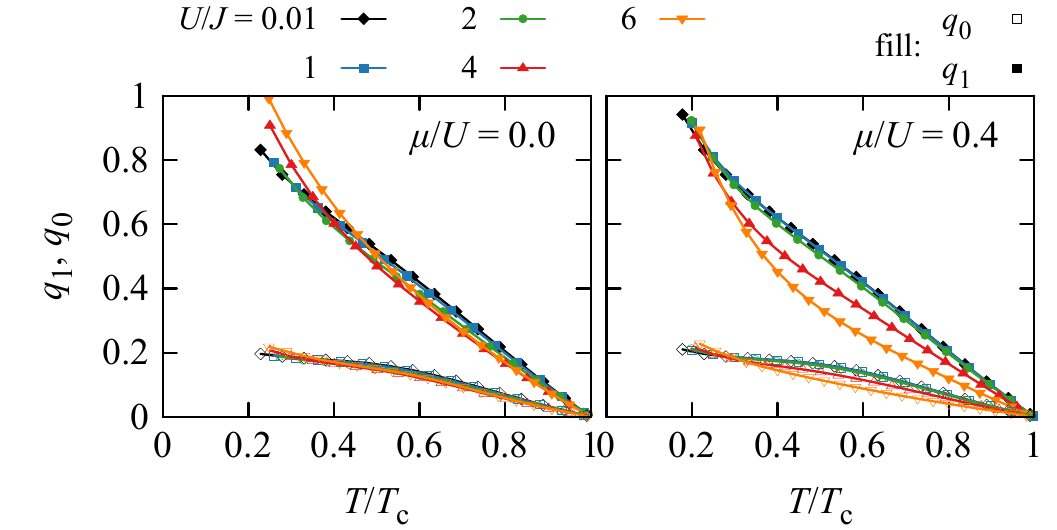}
	\caption{Temperature dependence of order parameters $q_{1}$ (full symbols) and $q_{0}$ (open symbols)
		at various values of the chemical potential $\mu/U$ (panels) and inverse disorder $U/J$ (point shapes),
		as marked on the plots.
		Lines are to guide the eye only.}
	\label{fig:q}
\end{figure}

{\it Glass order parameters.}---
We first study the 1-RSB composition of the $\Qaa$ matrix in the glass phase.
It is parametrized by the order parameters $q_{1}$ and $q_{0}$,
as well as the parameter $m$ that controls the split between the former two.
The order parameter $\QEA$ is defined as $\max(\Qaa)$~\cite{Parisi1979_PRL43},
which in 1-RSB equals $q_{1}$.
We study the dependence of these parameters on the reduced temperature $T/\Tc$.
First, we focus on $m$, depicted in Fig.~\ref{fig:m} at several values of $\mu/U$ and $U/J$.
The behavior is qualitatively similar across all parameter sets 
and is lobe-shaped. The curves approach zero near $T = \Tc$, exhibit a clear maximum at around $T/\Tc \approx 0.6$,
and again trend towards zero at $T = 0$,
which, however, is not reachable due to the increasing Trotter error~\cite{Suzuki1985_PLA113}.
This is consistent with the behavior of $m$ in spin-glasses~\cite{Parisi1980_JPAMG13,Buettner1990_PRB42}.
At fixed $\mu/U$, as $U/J$ increases, indicating that the system exhibits more local density correlations,
the values of $m$ decrease, and the curve becomes more asymmetric.
This indicates an interplay between density fluctuations, suppressed by interactions, and phase freezing, reflected in changing RSB structure.
A similar scenario was present in a quantum spin glass~\cite{Buettner1990_PRB42},
where the driving factor was the transverse field $\varDelta$, responsible for introducing quantumness to the system.
We might conclude that the nature of the studied glass phase approaches that of a spin glass as $U/J\to 0$.
Additionally, at $\mu/U = 0$ the curve saturates as $U/J$ increases, as the system is already locked to degenerate sectors of 0 and 1 particles per site,
while at $\mu/U = 0.4$ the decrease continues, since the system can access more sectors.
Above $U/J \gtrapprox 8$, there is no glass at any temperature~\cite{Piekarska2018_PRL120}.

Next, we turn our attention to the glass order parameters $q_{1}$ and $q_{0}$.
Their dependence as a function of $T/\Tc$ for the same set of $\mu/U$ and $U/J$ is shown in Fig.~\ref{fig:q}.
The general qualitative behavior mimics that of spin glasses again:
both parameters start at zero at $T = \Tc$ and increase as the temperature decreases,
with $q_{0}<q_{1}$.
Here, at $\mu/U = 0.4$, where interaction effects are more pronounced, we again observe their interplay with glassiness. Both $\qi=\mathcal{Q}_{\mathrm{EA}}$ and $\qi-\qo$ decrease, the former quantifying the strength of phase freezing and the latter the degree of RSB, revealing a competition between density fluctuations and off-diagonal glassiness.

\begin{figure}[htb]
	\includegraphics[width=\columnwidth]{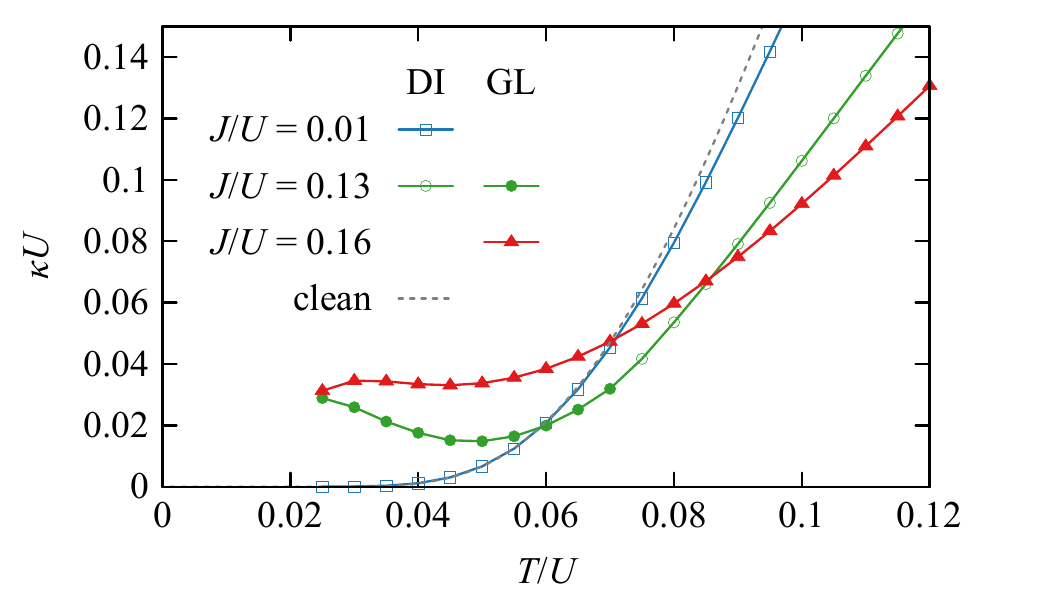}
	\caption{Compressibility $\kappa U$
	as a function of temperature $\kB T/U$
	for $J/U = 0.01, 0.13, 0.16$.
	Full symbols indicate the glassy phase, while open symbols stand for the disordered one.
    The dashed line shows a prediction of $\beta\exp(-\beta \mu)$ for a clean system.
	Lines are to guide the eye only.}
	\label{fig:compr}
\end{figure}

{\it Compressibility.}---
In the limit of no disorder, $J/U\to 0$, we recover the clean system behavior at zero hopping.
In this regime, only the disordered phase survives.
At sufficiently low temperatures, this phase is the incompressible Mott insulator phase.
As the disorder increases, the glass phase emerges.
In Fig.~\ref{fig:compr}, we plot the compressibility $\kappa = \partial \avg{n} / \partial \mu$
as a function of $T/U$, at fixed $\mu/U = 0.4$, and for three values of $J/U$.
At nearly zero disorder, $J/U = 0.01$, the system remains in the disordered phase throughout the plotted range.
The compressibility vanishes as $T\to 0$, as expected for the (almost) clean system.
The temperature dependence is also very close to that of the clean system $\kappa(\beta) = \beta \exp(-\beta \mu / U)$ with a gap $\mu/U$.
At the highest disorder considered, $J/U = 0.16$, the system remains in the glassy phase.
The compressibility dependence is significantly different,
with a low-$T$ plateau, indicating non-zero compressibility at $T\to 0$.
For intermediate disorder, $J/U = 0.13$,
the system crosses the phase transition at $T/U\approx 0.7$,
which is accompanied by a pronounced change in the trend.
The high-$T$ part of the curve seems to be decreasing to $\kappa = 0$,
but with the onset of glassiness, the curve turns upwards, and approaches a $\kappa\neq0$ plateau as $T\to 0$.
With this, we find a direct correspondence between compressibility and glassiness, and thus RSB.
It is nontrivial, as $\kappa=\beta(\avg{n^2}-\avg{n}^2)$ probes density fluctuations (diagonal sector of the Fock space), while glassiness involves the freezing of the bosonic phase (off-diagonal sector).
It has been, however, signaled by the interplay of interactions and glassiness in Figs.~\ref{fig:m} and \ref{fig:q}.
 
\begin{figure}[htb]
	\includegraphics[width=\columnwidth]{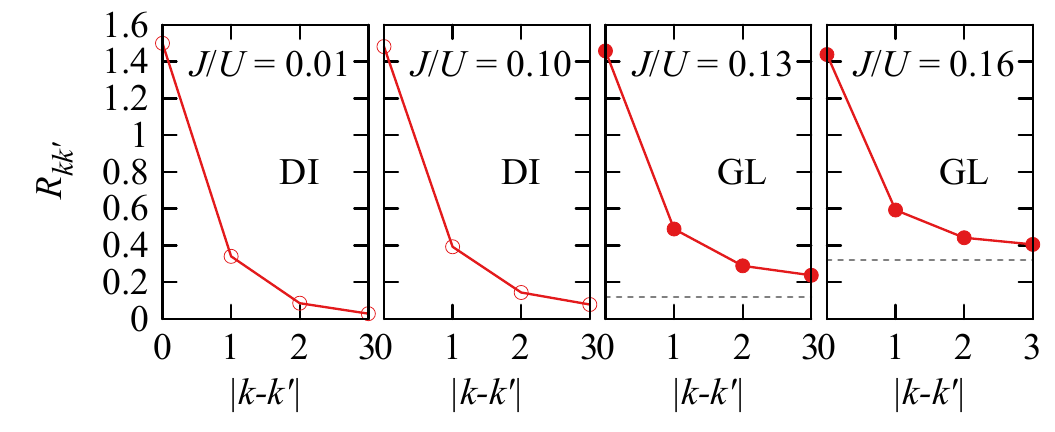}
	\caption{Self-correlations $\Rkk$ as a function of $|k-k'|$
		at $T/U = 0.05$ and varying values of $J/U$, as marked on the plots.
		The lower two values of $J/U$ lie within the disordered phase,
		while the higher two are in the glassy phase.
		In the glass, $\QEA$ is marked with dashed lines.
		Lines are to guide the eye only.}
	\label{fig:Rkk}
\end{figure}

{\it Self-interactions.}---
The dynamic self-interactions $\Rkk$ are a discrete equivalent of the autocorrelation function $\avg{a(\tau)a^\dagger(0)}$
depending only on the difference $|k-k'|$,
which is a counterpart of the imaginary Matsubara time $\tau$.
At $T=0$, the asymptote of the correlation function is expected to coincide with $\QEA$~\cite{book_Sachdev2023}, $[ \langle a^\dagger(\tau)a(0)\rangle ] = [ |\langle a\rangle|^2 ] + [ \langle \delta a^\dagger(\tau)\delta a(0)\rangle ] \xrightarrow{\tau\to\infty} [ |\langle a\rangle|^2 ] = \mathcal{Q}_{\mathrm{EA}}$.
In the discrete case, the asymptote can be recovered at $|k-k'| = M/2$.
In Fig.~\ref{fig:Rkk}, we present $\Rkk$ as a function of $|k-k'|$ at four values of $J/U$,
two in the disordered phase and two in the glassy phase.
For the latter two, we also plot the corresponding value of $\QEA$.
As expected, the correlations decay to the value of $\QEA$,
but do not fully saturate at finite temperature, due to an upper bound on imaginary-time $\tau\le\beta$.
The persistence of low-frequency (long-time) correlations indicates the presence of memory in the system, associated with the frozen bosonic phases.

{\it Discussion.}---
We have studied a phase transition to a non-ergodic glass phase in a system of interacting bosons with random hopping, employing the 1-RSB approach, which is essential for obtaining meaningful results in the RSB phase.
The phase involves random freezing of bosonic phases characterized by the EA order parameter.
We have analyzed the behavior of the order parameters of the system
and found limiting connections to spin glasses, quantum spin glasses, and clean Bose-Hubbard systems.
We have shown a straightforward distinction between the Mott-insulator and glass phases based on the zero-temperature limit of compressibility that is finite in the glass phase.
This central result is not obvious, as it relates a global density response to the freezing of local phases, involving operators that are diagonal and off-diagonal in the Fock basis, respectively.
Compressibility is routinely measured in optical lattice experiments,
typically aimed at distinguishing between Mott insulator and superfluid phases.
In contrast, conventional methods for identifying glassiness are more complex,
as they require dealing with real replicas or prohibitively long measurement times.
Thus, our results provide an experimentally well-established method for identifying glassiness.


\begin{widetext}
\section*{}
\renewcommand{\theequation}{A\arabic{equation}}
\setcounter{equation}{0}
{\label{sec:rsbderiv}\it Appendix A: Derivations.---}
We plug the $\lkaka$ decomposition from Eqs.~\eqref{eq:decomp-a}-\eqref{eq:decomp-b} into the free energy in Eq.~\eqref{eq:frene},
with the goal of transforming the expressions in a way that permits taking the $n\to 0$ limit.
The effective Hamiltonian in Eq.~\eqref{eq:effham} contains three terms of similar structure.
We handle all of them in a similar way, for example
\begin{multline}
	\Skaka \Qaa \pka \pkkaa = \sum_{kk'} \lr[\Big]{
		q_{0} \sum_{\alpha\alpha'} \pka \pkkaa +
		(q_{1}-q_{0}) \sum_{b} \sum_{\alpha\alpha'\in b} \pka \pkkaa +
		(-q_{1}) \sum_{\alpha} \pka p_{\alpha}^{(k')}
	} \\
	= q_{0} \lr[\Big]{\Ska \pka}^{2} +
	  (q_{1}-q_{0}) \sum_{b} \lr[\Big]{\sum_{k}\sum_{\alpha\in b} \pka}^{2} +
	  (-q_{1}) \sum_{\alpha} \lr[\Big]{\sum_{k} \pka}^{2},
\end{multline}
where $b$ ($1\leq b\leq m$) numbers the blocks of the $\Qaa$ matrix
and $\alpha \in b$ means that index $\alpha$ belongs to block $b$,
or, mathematically, that $\ceil{(b-1)n/m}\leq\alpha\leq\ceil{bn/m}$.
Next, we group the terms by the character of their $\alpha$-dependence and rearrange them such that each term is a squared sum, for example
\begin{equation}
	q_{0} \lr[\Big]{\Ska \pka}^{2} + q_{0} \lr[\Big]{\Ska \qka}^{2} + 2w_{0} \lr[\Big]{\Ska \pka}\lr[\Big]{\Ska \qka} =
	(q_{0}-w_{0}) \lr[\Big]{\Ska \pka}^{2} + (q_{0}-w_{0}) \lr[\Big]{\Ska \qka}^{2} + w_{0} \lr[\Big]{\Ska \pka + \Ska \qka}^{2}.
\end{equation}
Finally, we use the Hubbard-Stratonovich transformation and introduce a set of fields that decouple these squared sums, for example
\begin{multline}\label{eq:ema:field}
	\exp \lr*{\frac{1}{2}\lr*{\JBM}^{2}\Lr[\bigg]{
		d_{0} \lr[\Big]{\sum_{\alpha} P_{\alpha}}^{2} +
		\sum_{b} d_{01} \lr[\Big]{\sum_{\alpha\in b} P_{\alpha}}^{2}
		-d_{1} \sum_{\alpha} P_{\alpha}^{2}
	}} \\
	= \frac{1}{\sqrt{\pi}} \intid z \e^{-z^{2}} \exp \lr*{ \JBM z \sqrt{2d_{0}} \sum_{\alpha} P_{\alpha}}
		\prod_{b} \frac{1}{\sqrt{\pi}} \intid y_{b} \e^{-y_{b}^{2}} \exp \lr*{ \JBM y_{b} \sqrt{2d_{01}} \sum_{\alpha\in b} P_{\alpha} }
		\exp \lr*{ -\frac{1}{2}\lr*{\JBM}^{2}d_{1} \sum_{\alpha} P_{\alpha}^{2} } \\
	= \DD z \prod_{b} \DD y_{b} \prod_{\alpha \in b} \exp \lr*{
		\JBM z \sqrt{2d_{0}} P_{\alpha} +
		\JBM y_{b} \sqrt{2d_{01}} P_{\alpha} -
		\frac{1}{2}\lr*{\JBM}^{2}d_{1} \sum_{\alpha} P_{\alpha}^{2}
	},
\end{multline}
where we have introduced $P_{\alpha} \equiv \sum_{k} \pka$,  $d_{01} \equiv q_{1}-q_{0}-w_{1}+w_{0}$, $d_{0} = q_{0}-q_{1}$ and $d_{1} = q_{1}-w_{1}$.
This form is now suitable to take the $n\to 0$ limit using
\begin{multline}
	\DD z_{P} \DD z_{Q} \DD z_{B} \prod_{b} \DD y_{Pb} \DD y_{Qb} \DD y_{Bb} \prod_{\alpha\in b}
		\Tr \exp \lr*{ -\beta \lr* {H_{Pb\alpha}+H_{Qb\alpha}+H_{Bb\alpha}+H_0}}\\
	= \DD z_{P} \DD z_{Q} \DD z_{B} \Lr* { \DD y_{P} \DD y_{Q} \DD y_{B} \lr*{\Tr \e^{-\beta H}}^{m}}^{\frac{n}{m}},
\end{multline}
where $-\beta H_{Pb\alpha}$ is the exponent from the final expression in Eq.~\eqref{eq:ema:field},
$H_{Qb\alpha}$ and $H_{Bb\alpha}$ are analogous expressions stemming from sums of $\qka$ and $\pka+\qka$ respectively,
$H_{0}$ is the part that does not depend on $\alpha$ and $-\beta H$ is the final effective Hamiltonian that can be found in Eq.~\eqref{eq:finalham}.
\end{widetext}

\renewcommand{\theequation}{B\arabic{equation}}
\setcounter{equation}{0}
{\label{sec:numerics}\it Appendix B: Numerical calculations.---}
The self-consistent equations~\eqref{eq:selfcon-a}-\eqref{eq:selfcon-f} are solved numerically for fixed values of $T$, $U$, $\mu$, and $J$:
We assume starting values for the left-hand-side variables
and, repeatedly, use these to calculate the right-hand-side expressions,
which are then used as new values of the variables in question.
The calculation of the right-hand-side expressions presents two numerical challenges that must be optimized:
the computation of integrals $\DD z$ and $\DD y$ and the calculation of the partition function $Z$.

Each of the symbols $\DD z$, $\DD y$ represents a triple integral, which requires the calculation of six nested integrals in total.
This puts the main focus on reducing the number of evaluation points for each integral.
The integrals $\DD z$ and $\DD y$ can be efficiently approximated using the Gaussian quadrature method~\cite{Greenwood1948_BAMS54}.
We conducted a convergence study of various observables as a function of the number of integration points
and found that subsequent integers alternate in approximating the observables from above and below,
with improving accuracy.
We observed that the average of the $n$- and $n+1$-point integrals
provides accuracy similar to a $2n$-point calculation.
Thus, the results shown in the paper are calculated using $6$ and $7$ integration points and then averaged,
resulting in errors on the order of $1\%$.
It is often beneficial to treat a multiple integral as a whole,
rather than as a sequence of nested integrals, as we are doing here.
However, the Gaussian quadrature method offers no straightforward generalization to multiple dimensions.
Nevertheless, we examined several ideas in this regard, but none proved fruitful.

To calculate the partition function, we need to sum over all possible sets of variables $\{p_{1}, p_{2}, \ldots, p_{M}, q_{1}, q_{2}, \ldots, q_{M}\}$.
The sets of eigenvalues of the operators $\hat{P}$ and $\hat{Q}$ (or, equivalently, $\hat{n}$) are infinite,
necessitating the truncation of the Hilbert space.
The natural truncation is to limit the $\hat{n}$ basis
and remain within low values of the chemical potential $\mu$, as this reduction has physical meaning.
However, the number of summands in the partition function remains exponential,
which makes it very large, even when considering only a few particles per site.
One might attempt to use the Monte Carlo summation method.
However, unlike in the usual case, the summands might be complex or negative.
This gives rise to the severe sign problem~\cite{Piekarska2019_APPA135},
which renders the Monte Carlo method unsuitable for this calculation.
Thus, we must resort to direct summation.
We settle on a basis of up to $2$ particles per site,
allowing us to obtain results at $\mu/U \leq 1$ with reasonable accuracy.
This truncation leaves us with $9^{M}$ summands in the partition function.

\begin{acknowledgments}
{\it Acknowledgments.}---
Calculations have been partially carried out using resources provided by Wroclaw Centre for Networking and Supercomputing (\url{http://wcss.pl}), grant No. 449.
\end{acknowledgments}

\bibliography{all}
\bibliographystyle{apsrev4-2}

\end{document}